\documentclass[11pt,a4paper]{article}
\usepackage{amsmath,amssymb,amsthm,cite,latexsym,mathrsfs}
\usepackage[all,2cell,dvips]{xy}

\newcommand{\lbl}[1]{\label{eq:#1}}
\newcommand{\rf}[1]{(\ref{eq:#1})}
\newcommand{\nn}{\nonumber}

\def\eg{{\em e.g. }}

\renewcommand{\l}{\left}
\renewcommand{\r}{\right}

\newcommand{\Llra}{\ \Longleftrightarrow\ }

\renewcommand{\b}[1]{\mathbb{#1}}
\newcommand{\f}[1]{\mathfrak{#1}}

\newcommand{\sm}[2]{\textstyle{\frac{#1}{#2}}\displaystyle}
\newcommand{\shalf}{\textstyle{\frac{1}{2}}\displaystyle}

\newcommand{\prt}{\partial}

\begin{document}


\begin{titlepage}

\renewcommand{\thefootnote}{\fnsymbol{footnote}}

\begin{center}
{\LARGE

\indent

\indent

{\bf Polyakov soldering and second order frames~:}\\
{\bf the role of the Cartan connection}\\[2mm]
}

\indent


\vskip 1cm

{\Large
Serge Lazzarini\,\footnote{e-mail\,: \texttt{sel@cpt.univ-mrs.fr}} 
and Carina Tidei \footnote{e-mail\,: \texttt{tidei@cpt.univ-mrs.fr}~;
  allocataire de recherche MESR.} }\\[1cm] 

{\large
{\sc Centre de Physique Th\'eorique}
\footnote{Unit\'e Mixte de Recherche
(UMR 6207) du CNRS et des Universit\'es Aix-Marseille I, Aix-Marseille II et
de l'Universit\'e du Sud Toulon-Var. Unit\'e affili\'ee \`a la
FRUMAM F\'ed\'eration de Recherche 2291.}\\ 
CNRS Luminy -- case postale 907,\\ 
 F--13288 Marseille Cedex 9, France.}
\end{center}

\vskip 2cm

{\large
\centerline{\bf Abstract}

  The so-called `{\em soldering}' procedure performed by A.M.~Polyakov
  in \cite{Pol90} for a $SL(2,\mathbb{R})$-gauge theory is
  geometrically explained in terms of a Cartan connection on second
  order frames of the projective space $\b{R}$P$^1$.  
The relationship between a Cartan connection and the usual (Ehresmann)
connection on a principal bundle allows to gain an appropriate insight
into the derivation of the genuine `{\em 
  diffeomorphisms out of gauge transformations}' given by Polyakov himself.
}

\vfill

\noindent
PACS-2006 number :
02.40.Dr Euclidean and projective geometries,
02.40.Hw Classical differential geometry,
11.15.-q Gauge field theories, 
11.25.Hf Conformal field theory, algebraic structures.

\noindent
MSC-2000 number :
57R25 Vector fields, frame fields.

\medskip


\indent

\noindent
CPT-P013-2008

\end{titlepage}

\renewcommand{\thefootnote}{\arabic{footnote}}
\setcounter{footnote}{0}

\pagestyle{plain}

\section{Introduction}

\indent

More than fifteen years ago, in a paper by A.M. Polyakov \cite{Pol90},
diffeomorphism transformations  
for the 2-d conformal geometry were explicitly derived by a `\emph{soldering
procedure}' from a partial gauge fixing of two components out of three of a 
chirally split $SL(2,\mathbb{R})$-connection in the light-cone formulation.
In doing so, Polyakov ended with a residual
gauge transformation which exactly reproduced the Virasoro group
action on the effective energy-momentum tensor.
This striking result was raised by
Polyakov himself as `{\em a 
geometrical surprise}' and in addition some of the gauge parameters
were noticed to be gauge field dependent.

Subsequently, some work \eg \cite{Ver90,OSSvN92} directly referred to  
the Polyakov partial gauge fixing, which we shall call in the sequel as
the `{\em Polyakov soldering}'. It has to be said that in \cite{OSSvN92} this
intriguing result obtained by Polyakov was also emphasized as such.

In the present paper a proper differential geometrical framework is
proposed in order to 
explain this `{\em geometrical surprise}' \cite{Pol90,OSSvN92}. It is
essentially grounded on the use of the Cartan connection on the second
order frame 
bundle \cite{Kob72} and all the differential algebraic setup which goes with.

\medskip

In order to have a concise and efficient geometrical writing, we shall use the
BRS differential algebra for treating the gauge symmetry aspects.

One of the main ingredients will be the so-called solder forms on
frame bundles \cite{Kob61},  
well known objects in mathematics and 
intuitively figured out by Polyakov under the phrasing `{\em soldering}'
procedure. 

\section{Second order frames}

Let us briefly introduce the notion of the second order frame bundle.
Most of the time, we shall adopt the viewpoint that physics often requires,
the use of local expressions over a manifold. 
In other words, local comparison of a $n$-dimensional manifold $M$
with $\b{R}^n$ will be of constant use. 

According to \cite{Kob72,Sau89}, consider the principal bundle $F_2M$ of
second order frames, or shortly $2$-frames 
\footnote{It can be shown that they are torsion free linear frames
over the linear frame bundle itself \cite{Gra06}}, over $M$.
The elements of $F_2M$ are 2-jets $j_2(f)(0)$ of local
diffeomorphisms $f$ from a neighborhood of 0 in $\b{R}^n$ to $M$ such
that $f(0)=x$. Given local coordinate systems $\{x^\mu\}_{\mu=1}^n$ on
$M$ and $\{u^a\}_{a=1}^n$ on $\b{R}^n$, $F_2M$ is equipped with local
coordinates\footnote{Usual notation for gravitation will be constantly
  used \cite{EGH80}.}
$(x^\mu, e^\mu_{~a}, e^\mu_{~ab})$ corresponding to the
successive derivatives of $f$ at 0. Hence $e^\mu_{~a} \neq
0$ and $e^\mu_{~ab}=e^\mu_{~ba}$. 
When the composition of maps makes sense, the jet product
is given by 
the chain rule derivation $j_2(f)\cdot j_2(f') = j_2(f\circ f')$.

Accordingly 2-frame fields are sections of the bundle $F_2M$ over $M$,
$e_2(x) = \l(x^\mu, e^\mu_{~a}(x), e^\mu_{~ab}(x)\r)$ with inverse
given by
$e_2^{~-1}(x) = \l(x^\mu, e^a_{~\mu}(x), e^a_{~\mu\nu}(x)\r)$ where 
\begin{eqnarray}
e^a_{~\mu}e^\mu_{~b} = \delta^a_{b}, \qquad\qquad e^a_{~\mu\nu} = -
e^\lambda_{~bc}e^b_{~\mu}e^c_{~\nu}e^a_{~\lambda}\,. 
\lbl{invrep}
\end{eqnarray}
In the $u$-coordinates, a 2-frame $e_2$ at $x\in M$ has the polynomial
representative $f$, 
\begin{eqnarray*}
f^\mu(u) = x^\mu + e^\mu_{~a} u^a +
e^\mu_{~ab}u^au^b, \qquad u \in \b{R}^n\ , 
\end{eqnarray*}
such that $e_2 = j_2(f)(0)$,
and the coordinates $e^\mu_{~ab}$ and $e^\mu_{~a}$
are independent variables. Moreover, note that the 2-frame at $x$
associated to the local $x$-coordinates themselves are obtained as
2-jets of translations 
$(x^\mu,\delta^\mu_{~a},0) =j_2(u\mapsto x+u)(0)$. This induces a {\em
  natural} 2-frame at $x$ \cite{CCL99}. 

The structure group of $F_2M$ is the so-called differential group of order two
$G_2$ \cite{EMS91} locally given by the 2-jets $g_2=(g^a_{~a'},g^a_{~a'b'})$
at $0$ of diffeomorphisms $g$ of $\b{R}^n$ fixing the
origin\footnote{The differential group is the little group relative to
  the origin.}
$0$. 
The right action given by jet product $e'_2 = e_2\cdot g_2 =
j_2(f\circ g)(0)$ reads 
\begin{eqnarray*}
f (g (0)) = f(0) = x,\qquad e^\mu_{~a'} = e^\mu_{~a}\,g^a_{~a'},
\qquad e^\mu_{~a'b'}=e^\mu_{~ab}\,g^a_{~a'}g^b_{~b'} + e^\mu_{~a}\,g^a_{~a'b'}. 
\end{eqnarray*}
It is worthwhile to note the semi-direct product decomposition
of $G_2 = GL_0 \ltimes GL_1$ with respect to the jet product, namely
$g_2=(g^a_{~b'},0)\cdot(\delta^{b'}_{~a'},g^{b'}_{~b}g^b_{~a'b'})$, with
$(g^a_{~a'}) \in GL_0 := GL(n,\b{R})$ 
\footnote{The motivation for the lower indices attached to $GL$ will
become clear below in the main text.}. 

To the local $x$-coordinates on $M$, there corresponds a {\em natural
gauge} in which the local coordinates of any
2-frame field  $e_2: x\mapsto e_2(x)\in G_2$ can be considered as an
element of the gauge 
group, thanks to the pointwise identification
$\l(e^\mu_{~a}(x),e^\mu_{~ab}(x)\r) =
\l(\delta^\mu_{~c},0\r)\cdot\l(g^c_{~a}(x),g^c_{~ab}(x)\r)$.  

Let us now introduce a family of solder 1-forms on $F_2M$ (also called
canonical forms) which are invariant under 
diffeomorphisms of $M$ by taking the Maurer-Cartan like 1-form ``$f^{-1}
\circ d f$" in powers of $u\in \b{R}^n$ \cite{Per01},
\begin{eqnarray*}
	(f^{-1} \circ d f)^a (u) = \theta^a + \theta^a_{~b}\, u^b +
        \sm{1}{2} \theta^a_{~bc}u^bu^c + \cdots\,.
\end{eqnarray*}
Then, in terms of the local coordinates
$(x^\mu,e^\mu_{~a},e^\mu_{~ab})$ on $F_2M$ the solder 1-forms read
\footnote{$e^a_{~\mu}$ denotes the usual tetrad \cite{EGH80}.}  
\begin{eqnarray}
  \theta^a = e^a_{~\mu} dx^\mu,\qquad \theta^a_{~b} = e^a_{~\mu} d e^\mu_{~b} +
  e^a_{~\mu\nu} e^\mu_{~b} dx^\nu\,.
\lbl{solde}
\end{eqnarray}
In particular, they fulfil the torsion free condition $d\theta^a + \theta^a_{~b}
\wedge \theta^b=0$.

\section{Cartan connection on second order frames}

Adding translations denoted as $g^a\in GL_{-1}\simeq\b{R}^n$, consider
now the structure group $G_2$ as a Lie subgroup of the Lie group
$G:=GL_{-1}\times G_2$. Denoting $\f{gl}_k :=\textrm{Lie~}GL_k, k = 0,
\pm1$, the Lie algebra $\f{g}$ of $G$ admits the decomposition $\f{g}
= \f{gl}_{-1} \oplus \f{gl}_0 \oplus \f{gl}_1$. In the sequel, we set
$\f{g}_2:= \textrm{Lie~}G_2 = \f{gl}_0\oplus\f{gl}_1$.   
Since $G$ consists of 2-jets at $0$ of diffeomorphisms
$g_t(u) = u+tX(u)$ of $\b{R}^n$, with $g_0=id_{\b{R}^n}$, $X(u)=\l(X^a+X^a_{~b}u^b+X^a_{~bc}u^bu^c+\cdots\r)\partial_a$,
its Lie algebra $\f{g}$ consists of tangent vector fields at the
identity $\bold{1}:=j_2(id_{\b{R}^n})(0)= \l(\delta^a_{~b},0\r)$,
given by the 2-jet at $0$ of the vector field $X(u)$ on $\b{R}^n$,
\[X_2 := j_2(X)(0) = \l.\frac{d}{dt}\r|_{t=0}j_2(g_t)(0) =
(X^a,X^a_{~b},X^a_{~bc}) . \]  
The algebraic bracket is then defined to be \emph{minus} the Lie
bracket of vector fields \cite{Mil84}, 
\begin{eqnarray}
&& [X,Y]^a = X^a_{~b} Y^b - Y^a_{~b} X^b\nn\\
&& [X,Y]^a_{~b} = X^a_{~c} Y^c_{~b} + X^a_{~bc} Y^c - (X
\leftrightarrow Y)
\lbl{algbrac}\\
&& [X,Y]^a_{~bc} = X^a_{~d} Y^d_{~bc} + X^a_{~dc} Y^d_{~b} + X^a_{~bd}
Y^d_{~c} - (X \leftrightarrow Y).\nn
\end{eqnarray}
Then, $[\f{gl}_{-1},\f{gl}_{-1}]=0$ for the translation part and
$[\f{gl}_k,\f{gl}_{\ell}]\subset \f{gl}_{k+\ell}, k,\ell=0,\pm1$. So,
the Lie algebra $\f{g}$ turns out to be a graded Lie algebra
\cite{Kob72} with respect to the dilatation generator $u^a\partial_a$.  

Moreover, considering the  differential group of order three $G_3
\ni~g_3~=~\l(g^{a'}_{~a},g^{a'}_{~ab},g^{a'}_{~abc}\r)$, an adjoint
type action on $\f{g}$ can be defined by \cite{Kob61} 
\begin{eqnarray*}
\textrm{Ad}(g_3)\,X_2 = \l.\frac{d}{dt}\r|_{t=0}j_2(g'\circ g_t\circ g'^{-1})(0) = \l(X^{a'}, X^{a'}_{~b'}, X^{a'}_{~b'c'}\r)\,,
\end{eqnarray*}
where the transformed components are given by chain rule derivations as,
\begin{eqnarray}
X^{a'} &=& g^{a'}_{~a}X^a\ ,\nn\\[2mm]
X^{a'}_{~b'} &=& g^{a'}_{~ab}X^b g^a_{~b'} +
g^{a'}_{~a}X^a_{~b}g^b_{~b'}\ ,
\lbl{adj}\\[2mm]
X^{a'}_{~b'c'} &=& g^{a'}_{~abc}X^cg^b_{~b'}g^a_{~c'} +
g^{a'}_{~ab}X^b_{~c}g^c_{~b'}g^a_{~c'} +
g^{a'}_{~ac}g^c_{~b'}X^a_{~b}g^b_{~c'}\nn\\[2mm] 
&&+\, g^{a'}_{~a}X^a_{~bc}g^c_{~b'}g^b_{~c'} +
g^{a'}_{~ab}X^bg^a_{~b'c'} + g^{a'}_{~a}X^a_{~b}g^b_{~b'c'}\,. \nn
\end{eqnarray}

Now, we are in position to introduce a Cartan connection on $F_2M$ as
a $\f{g}$-valued 1-form satisfying: 
\begin{itemize}
\item $\forall e_2 \in F_2M$, $\omega_{|_{e_2}} : T_{e_2} \l(F_2M\r) \to
  \f{g} $ is a linear 
isomorphism (absolute parallelism),
\item $\forall g_2 \in G_2 \subset G_3$, $R_{g_2}^*\omega =
  \textrm{Ad}({g_2}^{-1})\,\omega$, where $g_2 =
  \l(g^{a'}_{~a},g^{a'}_{~ab},0\r)$ ($G_2$-equivariance),  
\item $\forall \hat{X} \in V\l(F_2M\r)$, $\omega(\hat{X}) \in \f{g}_2$.
\end{itemize}

According to the grading of $\f{g}$, the Cartan connection $\omega$
can be chosen to be \cite{Kob72}
\begin{eqnarray*}
\omega := \omega_{-1} + \omega_0 + \omega_1 = \theta^a + \theta^a_{~b}
+ \omega^a_{~bc}\,, 
\end{eqnarray*}
where $\theta^a$ and  $\theta^a_{~b}$ are the solder forms on $F_2M$
which are respectively $\f{gl}_{-1}$-valued, and $\f{gl}_0$-valued
1-forms, and $\omega^a_{~bc}$ is a $\f{gl}_1$-valued 1-form on
$F_2M$. Its curvature is defined by
\begin{eqnarray*}
K = d\omega + \frac{1}{2}[\omega,\omega] = K_{-1} + K_0 + K_1
\end{eqnarray*}
where the bracket \rf{algbrac} has to be used. More explicitly one finds
the torsion free condition $K_{-1} = d\omega_{-1} +
\omega_0\wedge\omega_{-1} = 0$ 
and the (generalized) curvature $K_0 = d\omega_0 +
\sm{1}{2}[\omega_0,\omega_0] + [\omega_{-1},\omega_1]$.

\indent

For the sequel, it is of use to note that, from a Cartan connection on
$F_2M$ it is possible to recover the Yang-Mills context by
constructing a (Ehresmann) 
connection  on the principal $G$-bundle $F_2M \times_{G_2} G$, see
\cite{Sha97,Wis06}. Actually, at a point $(e_2,g) \in F_2M \times G$,
one can construct the $\f{g}$-valued $1$-form
$$
{\cal A}_{|(e_2,g)} = \textrm{Ad}(g^{-1}) (\pi_{F_2M})^* \omega +
(\pi_G)^* \Theta_G, 
$$
where $\Theta_G$ is the Maurer-Cartan form on $G$ and $\pi_{F_2M}$
(resp. $\pi_G$) is the canonical projection on $F_2M$ (resp. $G$). ${\cal
  A}$ turns out to be a connection $1$-form on the principal bundle
$F_2M \times_{G_2} G$ \cite{Sha97}. 

Since we are concerned with local expressions, consider the
local connection $1$-form $A=\sigma^*{\cal A}$ (gauge field on $M$)
obtained as pull-back of 
the connection $1$-form by a (local) section $\sigma(x)$ of
$F_2M\times_{G_2} G$. 
When we restrict ourselves to $F_2M\simeq F_2M\times_{G_2}
\{\bold{1}\}$, and take $\sigma(x) = 
(e_2(x),\bold{1})$, the gauge field reduces to
$A=e_2^*\omega$, namely the local expression of the Cartan connection on $M$.
As it will be shown, this is the gauge field considered by Polyakov.

As outlined in \cite{DV80}, let us first consider, in the natural gauge, the
``local expression" on 
$F_2M$ of the Cartan connection $\omega$ (a gauge like redefinition
which to some extent is field dependent) 
\begin{eqnarray}
\Gamma(e_2,\omega) = \textrm{Ad}\l(\ell(e_2)\r)\,\omega + e_2\cdot
de_2^{-1},
\lbl{redef}
\end{eqnarray}
where $\ell(e_2)=\l(e^\mu_{~a},e^\mu_{~ab},e^\mu_{~abc}\r)$ is the
necessary lift of an element $e_2\in G_2$ into $G_3$. It can be shown that
the local connection 1-form $\Gamma$ generalizes the Christoffel
symbols to second order frames \cite{Gra06}. Moreover by extending the proof
for an affine connection \cite{CCL99} it can be checked that the generalized
Christoffel symbols depend only on the $x$-coordinate .
A direct computation~\footnote{In the course of
the computation using \rf{invrep} and \rf{adj} it is easier to
consider $e_2\cdot de_2^{-1} = - de_2 
\cdot e_2^{-1}$.} gives for the components of the torsion free
Cartan connection 
\begin{eqnarray}
\Gamma^\mu &=& e^\mu_{~a}\theta^a = dx^\mu\ ,\nn\\[2mm]
\Gamma^\mu_{~\nu} &=& e^\mu_{~a}\theta^a_{~b}e^b_{~\nu} + e^\mu_{~ab}\theta^ae^b_{~\nu} + e^\mu_{~a}de^a_{~\nu} = 0\ ,\lbl{gamma}\\[2mm]
\Gamma^\mu_{~\nu\rho} &=& e^\mu_{~abc}\theta^c e^b_{~\nu} e^a_{~\rho}
+ e^\mu_{~ac}\theta^ce^a_{~\nu\rho} +
e^\mu_{~ab}\theta^b_{~d}e^d_{~\nu} e^a_{~\rho} +
e^\mu_{~ac}e^c_{~\nu}\theta^a_{~b} e^b_{~\rho}\nn\\[2mm] 
&& +\, e^\mu_{~a}\theta^a_{~b} e^b_{~\nu\rho} +
e^\mu_{~a}\omega^a_{~bc}e^c_{~\nu} e^b_{~\rho} +  
e^\mu_{~a}de^a_{~\nu\rho} + e^\mu_{~ab}d(e^b_{~\nu} e^a_{~\rho}) =
\Gamma^\mu_{~\rho\nu}\ .\nn 
\end{eqnarray}

\section{Reduction to the projective case}

Having in mind the implementation of a $SL(2,\b{R})$ gauge symmetry,
we reduce right away the structure group $G = GL_{-1}\times G_2$ to
$SL(2,\b{R}) = SL_{-1}\times \l(SL_0 \ltimes SL_1\r)$ with the parametrization 
\begin{eqnarray*}
\b{R}\simeq SL_{-1} \ni \l(\begin{matrix}1&b\\0&1\end{matrix}\r),\quad
 SL_0\ltimes SL_1 \ni \l(\begin{matrix}a&0\\c & 1/a \end{matrix}\r) ,c \in
\b{R}, a \in \b{R}\setminus\{0\}\,. 
\end{eqnarray*}
Accordingly the Lie algebra $\f{sl}(2,\b{R})$ inherits the graded
splitting $\f{sl}_{-1} \oplus \f{sl}_0 \oplus \f{sl}_1$.

This reduction amounts to restricting ourselves to the
$1$-dimensional (real)
projective space ${M=\b{R}P^1}$ with local coordinate $x$. Let $u$
denote the local coordinate on $\b{R}$. The bundle
$P_2$ of projective 2-frames $\l(x,e^x_{~u},e^x_{~uu}\r)$ is a
principal sub-bundle of  
$F_2M$ with the reduced structure group $SL_0 \ltimes SL_1 \subset G_2$ (the
so-called $G$-structure \cite{Kob72}). More precisely, elements $g_2$
of the structure group are 2-jets at $u=0$ of the
projective maps fixing $u=0$ given by
\begin{eqnarray*}
 g(u) = \frac{a u }{cu+1/a} \Llra \binom{g(u)}{1} = \begin{pmatrix} a
   & 0 \\ c & 1/a \\ \end{pmatrix} 
  \binom{u}{1} .
\end{eqnarray*}
In particular, the identification of $SL_0 \ltimes SL_1$ as subgroup of $G_2$ is made through
\begin{eqnarray*}
	\begin{pmatrix} \l(g^{u'}_{~u}\r)^{1/2} & 0 \\ - \sm{1}{2}\,
          g^{u'}_{~uu}\l(g^{u'}_{~u}\r)^{-3/2} &
          \l(g^{u'}_{~u}\r)^{-1/2} \\ \end{pmatrix}. 
\end{eqnarray*}
The Cartan connection $\omega$ on $F_2M$ restricts to $P_2$ as the
$\f{sl}(2,\b{R})$-valued 1-form whose local expression (see
\rf{solde}) reads 
\begin{eqnarray}
\omega_{-1} = \theta^u =e^u_{~x}\,dx,\qquad \omega_0 = \theta^u_{~u} =
e^x_{~u} e^u_{~xx}\,dx + 
e^u_{~x}\,de^x_{~u},\qquad \omega_1=\omega^u_{~uu}\ . 
\lbl{solde'}
\end{eqnarray}
In the natural gauge, a projective 2-frame field $e_2(x) = j_2(f)(0)$ with
$f(u) = x+g(u)$ is viewed as an element $G_2$. Moreover
it is well known that projective transformations $y=f(u)$ are solutions
of the third order differential equation $y''' = \sm{3}{2} (y'')^2/y'$.
This induces a relation between the local coordinates of the whole
projective frame  $j_3(f)(0)=\l(e^x_{~u},e^x_{~uu},e^x_{~uuu}\r) \in G_3$, where 
the third order jet has the unique form
\begin{eqnarray}
e^x_{~uuu} = \sm{3}{2}\l(e^x_{~uu}\r)^2 e^u_{~x}.
\lbl{Schwarz}
\end{eqnarray}
The relationship \rf{Schwarz} gives rise to the unique lift
$\ell(e_2)$ of the projective 2-frames 
into $G_3$ dictated by the projective structure itself on $\b{R}P^1$.

By restricting the local expression \rf{redef} to the projective case,
and taking into account the expressions \rf{invrep}, \rf{solde'} and
\rf{Schwarz} the components \rf{gamma} of the Cartan connection 1-form on
$P_2$ read~: 
\begin{eqnarray}
\Gamma^x &=& dx \ ,\nn\\
\Gamma^x_{~x} &=& 0 \ , \lbl{gamma_proj} \\
\Gamma^x_{xx} 
&=& e^u_{~x}\omega^u_{~uu} + d\l(e^u_{~xx}e^x_{~u}\r) -
\sm{1}{2}\l(e^u_{~xx}e^x_{~u}\r)^2dx \ ,\nn 
\end{eqnarray}
where the special combination $\chi_x:=e^u_{~xx}e^x_{~u}$ occurs in a Miura
like expression \cite{Mat90}.
Under a change of local coordinates $x \mapsto
x'$ on $M$, the frame coordinates transform as 
\[ e^x_{~u} \to e^{x'}_{~u} = \frac{dx'}{dx}\,e^x_{~u},\qquad 
e^{x}_{~uu} \to e^{x'}_{~uu} = \frac{dx'}{dx}\,e^{x}_{~uu} + \frac{d^2
  x'}{dx^2}\l(e^x_{~u}\r)^2\] 
while for the inverse 
\[e^u_{~x'} = \frac{dx}{dx'}\,e^u_{~x},\qquad e^u_{~x'x'} =
\l(\frac{dx}{dx'}\r)^2e^u_{~xx} -
\l(\frac{dx}{dx'}\r)^3 \frac{d^2 x'}{dx^2}\,e^u_{~x}.\] 
It can be checked, on the one hand, that $\chi_x$ behaves as an 
affine connection, and,
on the other hand, that the third component $\Gamma^x_{~xx}$ 
behaves as a projective connection, namely 
\begin{eqnarray}
\Gamma^{x}_{~xx} - \frac{dx'}{dx}\,\Gamma^{x'}_{~x'x'} =
\l(\frac{d^3 x'}{dx^3} \left/ \frac{dx'}{dx} \right. -
\frac{3}{2}\l(\frac{d^2 x'}{dx^2}\left/
  \frac{dx'}{dx}\right.\r)^2\r) dx \ ,
\end{eqnarray}
where the right hand side is recognized as the Schwarzian derivative.

So, locally on $P_2$ the connection $\Gamma$ as a
$\f{sl}(2,\b{R})$-valued 1-form is parametrized as
\begin{equation}
\Gamma = \begin{pmatrix} 0 & dx \\ \Gamma^x_{~xx} & 0 \\
\end{pmatrix}\, .
\lbl{projparam}	
\end{equation}
Given a projective 2-frame field $e_2(x)$, let us denote the local
representative 
of the Cartan connection on $M$ by $e_{2}^*\omega = (\theta^u_{~,x} + 
\theta^u_{~u,x} + \omega^u_{~uu,x})dx$.
Thus \rf{projparam} is pulled-back to $M$ as 
\begin{eqnarray}
  \begin{pmatrix} 0 &
 1 \\ \Gamma^x_{~xx,x}(x) & 0 \\ 
\end{pmatrix} dx\, , \quad \mbox{with } \Gamma^x_{~xx,x}(x) =
e^u_{~x}\omega^u_{~uu,x} + \prt_x \chi_x - \sm{1}{2}\l(\chi_x \r)^2. 
\lbl{projparam_loc}	
\end{eqnarray}
Hence, the local expression of the Cartan connection $\omega$ on the
projective 2-frame bundle gives rise directly to a `\emph{Polyakov
  soldering}' procedure but in an appropriate geometrical framework. 

\section{The BRS-structure}

Let us now turn to the infinitesimal gauge aspect related to the
$SL(2,\b{R})$ Yang-Mills counterpart of the Cartan connection.
The infinitesimal $SL(2,\b{R})$-gauge transformations can be recast in
the more powerful and elegant BRS (graded) differential algebra
\cite{BRS76,Sto77} by turning the gauge
parameters to Faddeev-Popov ghost fields~$\gamma$. 

The infinitesimal gauge transformation is usually written in terms of
a nilpotent $s$-operation as
\begin{eqnarray}
s \omega = - d\gamma - [\omega,\gamma],\qquad
s\gamma = - \shalf [\gamma,\gamma],\qquad s^2 =0,
\lbl{BRSalg} 
\end{eqnarray}
where the graded bracket is understood with respect to the de Rham
degree and the ghost number (or BRS grading). The Lie algebra content
of the graded bracket is given by algebraic bracket~\rf{algbrac}.
 With respect to the bigrading, one has 
the nilpotency properties $(d+s)^2=0$, namely $d^2=s^2=ds+sd=0$.
Recall that the nilpotent algebra \rf{BRSalg} can be compactly
encapsulated into one formula only, the so-called Russian formula for
the field strength (the curvature)
\begin{eqnarray}
	d\omega+\sm{1}{2}[\omega,\omega] = (d+s)(\omega+\gamma) +
        \sm{1}{2}[\omega+\gamma,\omega+\gamma], 
	\lbl{Russform}
\end{eqnarray}
where $\omega+\gamma$ acquires the status of algebraic connection \cite{DbV86}.

Let us introduce now the projective parametrization \cite{GGL95} as
the redefinition 
\footnote{Limiting ourselves to 1-frames 
$\mbox{Ad}(\ell(e_1))(\omega + \gamma) + e_1\cdot(d + s)e_1^{-1}$
where $\ell(e_1) = (e^\mu_{~a},0)\in G_2$
exactly yields the conformal parametrization given in \cite{GGL95}.}
\begin{eqnarray}
\Gamma + c = \mbox{Ad}(\ell(e_2))(\omega + \gamma) + e_2\cdot(d + s)e_2^{-1}\ ,
\lbl{proj_param}
\end{eqnarray}
where the ghost field is redefined by $c = \mbox{Ad}(\ell(e_2))\gamma +
e_2\cdot s e_2^{-1}$.  It can be checked that 
$$
s\Gamma = - dc - [\Gamma,c],\qquad sc = -\,\sm{1}{2}[c,c].
$$

Recalling that the local representative $e_{2}^*\omega$ of the Cartan
connection is a  
$\l(\f{sl}_{-1}\oplus \f{sl}_0\oplus\f{sl}_1\r)$-valued 1-form on $M$, an
obvious $\f{sl}(2,\b{R})$-ghost parameter can be chosen to be
\begin{eqnarray*}
	\gamma = \l(e_{2}^*\omega\r)(\xi) = \omega({e_{2}}_*\xi)
	\lbl{lift}
\end{eqnarray*} 
where $\xi = \xi^x \prt_x$ is the ghost vector field defined on $M$.
More explicitly the components of that particular gauge ghost read
\begin{eqnarray}
\gamma^u = \theta^u_{~,x}\xi^x = e^u_{~x}\xi^x,\quad
\gamma^u_{~u} = \theta^u_{~u,x}\xi^x = \l(e^u_{~xx}e^x_{~u} +
e^u_{~x}\partial_xe^x_{~u}\r)\xi^x,\quad  
\gamma^u_{~uu} = \omega^u_{~uu,x}\xi^x .
\lbl{obv}
\end{eqnarray}
Then, performing the BRS transformations
\rf{BRSalg} on the components of the Cartan connection, with respect
to that particular ghost parametrization, one gets 
\begin{eqnarray}
&&s\theta^u_{~,x} = \partial_x\gamma^u + \theta^u_{~u,x}\gamma^u -
\gamma^u_{~u}\theta^u_{~,x} =
\partial_x\l(\theta^u_{~,x}\xi^x\r)\ ,\nn\\[2mm] 
&&s\theta^u_{~u,x} = \partial_x\gamma^u_{~u} +
\omega^u_{~uu,x}\gamma^u - \gamma^u_{~uu}\theta^u_{~,x} =
\partial_x\l(\theta^u_{~u,x}\xi^x\r)\ , \lbl{stheta}\\[2mm]
&&s\omega^u_{~uu,x} = \partial_x\gamma^u_{~uu} +
\omega^u_{~uu,x}\gamma^u_{~u} - \gamma^u_{~uu}\theta^u_{~u,x} =
\partial_x \l(\omega^u_{~uu,x}\xi^x\r)\ ; \nn
\end{eqnarray}
these variations can be compactly gathered as $s\omega = -
d(\omega(\xi)) = \l(i_\xi d - d i_\xi \r) =: L_\xi \omega$, which is
nothing but the Lie derivative\footnote{Remind that $\xi$ carries a
  ghost degree and that here $M$ is one dimensional.} expressing the action of
diffeomorphisms on the Cartan connection 1-form.

Using the local expressions \rf{solde'} of $\theta^u$ and $\theta^u_{~u}$, the
variations \rf{stheta} infer the lift of the diffeomorphisms on 2-frame
fields as BRS transformations given by
\begin{eqnarray}
se^u_{~x} &=& s\theta^u_{~,x} = \partial_x\l(e^u_{~x}\xi^x\r)\nn\\[2mm]
se^u_{~xx} &=& e^u_{~x}\,s\theta^u_{~u,x} + e^x_{~u}e^u_{~xx}\,se^u_{~x} +
e^x_{~u}\partial_xe^u_{~x} se^u_{~x} -
\l(e^u_{~x}\r)^2\partial_x\l(se^x_{~u}\r) \lbl{BRS_rep}\\[2mm]
&=& \partial_xe^u_{~xx}\xi^x + 2e^u_{~xx}\partial_x\xi^x +
e^u_{~x}\partial_x^2\xi^x \nn
\end{eqnarray}
Accordingly, in each sector of the graded algebra $\f{sl}_{-1}\oplus
\f{sl}_0\oplus\f{sl}_1$ we are respectively left with the following ghosts
\begin{eqnarray}
c^x &=& e^x_{~u}\gamma^u = \xi^x,\nn\\[2mm]
c^x_{~x} &=& e^x_{~uu}e^u_{~x}\gamma^u + \gamma^u_{~u} +
e^x_{~u}se^u_{~x} = \partial_x \xi^x 
\lbl{proj_c}\\[2mm]  
c^x_{~xx} &=& \sm{1}{2}\l(e^u_{~xx}\r)^2\l(e^x_{~u}\r)^3\gamma^u +
e^x_{~uu}\l(e^u_{~x}\r)^2\gamma^u_{~u} + \gamma^u_{~uu}e^u_{~x} +
2e^x_{~uu}e^u_{~x}se^u_{~x} + e^x_{~u}se^u_{~xx}\nn\\[2mm] 
&=& \partial^2\xi^x + \l(\partial_x\chi_x -\sm{1}{2}\l(\chi_x \r)^2 +
e^u_{~x}\omega^u_{~uu,x}\r)\xi^x\ =\ \partial^2_x\xi^x +
\Gamma^x_{~xx,x}\xi^x \ ,\nn 
\end{eqnarray}
which are exactly the ghost version of the gauge parameters found
by Polyakov in \cite{Pol90}. 
With the help of \rf{BRS_rep} the BRS transformations on the
connection form $\Gamma$ \rf{gamma_proj} are computed to~be
\begin{eqnarray}
s\Gamma^x_{~,x} = 0, \qquad s\Gamma^x_{~x,x} = 0, \qquad
s\Gamma^x_{~xx,x} = \partial_x^3\xi^x +
\xi^x\partial_x\Gamma^x_{~xx,x} + 2\partial_x\xi^x \Gamma^x_{~xx,x}. 
\end{eqnarray}
These variations derive from the geometry, while the first two were
imposed in \cite{Pol90} as constraints in order to keep the gauge choice.
The third variation can be rewritten as,
\begin{eqnarray}
s\Gamma^x_{~xx,x} = \partial_x^3c^x + c^x\partial_x\Gamma^x_{~xx,x} +
2\partial_xc^x \Gamma^x_{~xx,x}, 
\end{eqnarray}
where $c^x$ is the ghost vector field with $sc^x = c^x\partial_x c^x$
and $s^2=0$.

As claimed in \cite{Pol90}, this residual BRS algebra exactly
 reproduces the well known Virasoro 
algebra action on the energy-momentum tensor suitably identified with
$\Gamma^x_{~xx,x}$. 


\section{Conclusions}

Cartan geometry provides a proper framework which amounts to explain
both  the `geometrical surprise' expressed by Polyakov and
the field dependence of the  constraint on the gauge parameters.  As
shown, on the one hand, it turns out that the specific Polyakov
partial gauge fixing is just the local expression \rf{gamma_proj}
of the  Cartan connection
for which the first two components with respect to a graded Lie algebra are
those of the solder 1-form on the second order frame  bundle, while the
untouched third component of the Cartan connection acquires a
geometrical status, namely that of a projective connection. 
Recall
that the latter shares exactly the same geometrical feature of the
effective energy-momentum tensor in 2D-conformal chiral theory.  The
approach has the advantage of shedding some light on the Polyakov
soldering which is the main basis of the so-called $W$-gravity
\cite{OSSvN92,GGL95}.  Incidentally, the framework of Cartan geometry
gives a significant improvement of some manipulations performed in
\cite{GGL95}.

On the other hand, the relationship between a Cartan connection and
the usual (Ehresmann) connection on a principal bundle  has allowed us
to gain some insight into the derivation of ``{\em
diffeomorphisms out of gauge transformations}" given by Polyakov in
\cite{Pol90}. This has been achieved by lifting the diffeomorphisms from
the base manifold to the second order frame bundle, see \rf{BRS_rep}. All 
geometrical manipulations were carried out in terms of projective
2-frame fields.

\indent

Furthermore, the use of Cartan connections has been somewhat put
aside in  spite of a few attempts in the past \cite{BD77} as well as the
use of second order frames \cite{HP78}. However there is a revival
interest in the use of the concept of Cartan connections, especially in
(conformal) gravity or in the study of PDE's, see for instance
\cite{GL99,FKN01,FKN02,FKNN02,KL03,NN03,KNN03,GKNP04,Wis06}. An
extensive study of the Cartan connections and the related geometries
has been also performed by the mathematical Czech school see \eg
\cite{Cap08} and references therein. In particular, a
possible route in relation with the so-called $W$-symmetries as
extended conformal symmetries could have been
opened on the mathematical side by the study of parabolic geometries
and their Cartan connections \eg \cite{CS00,Cap06}.

The Polyakov soldering provides another physical
situation in which 
Cartan connections appear naturally.  They allow to lift the action
of diffeomorphisms to higher order differential structures which ought
to be of interest for  physical models of
gravitation.  They offer also a proper 
differential geometrical framework which allows a generalization of
the Polyakov soldering \rf{gamma} to higher dimensions \cite{Gra06}, for
instance when third order frames can be expressed in terms of the lower
ones (this is the case for projective and conformal structures
\cite{Kob72}).

\indent

\paragraph{Acknowledgements.} We would like to thank T.~Sch\"ucker for
comments and for a careful reading of the manuscript.
We are indebted to one of the Referees for pointing out to us some
references about the mathematical Czech school.



\bibliographystyle{unsrt}

\end{document}